\documentclass[11pt,twoside]{article}
\usepackage{asp2006}

\usepackage{epsf}
\usepackage{lscape}

\usepackage{graphicx}
\usepackage{amssymb,bm,color,multirow}
\def\edcomment#1{\iffalse\marginpar{\raggedright\sl#1\/}\else\relax\fi}
\reversemarginpar

\begin{document}

\title{3D MHD Simulations of Disk
Accretion onto Magnetized Stars: Numerical Approach and Sample
Simulations}
\author{Marina M. Romanova\altaffilmark{1}, Alexander V. Koldoba\altaffilmark{2},
Galina V. Ustyugova\altaffilmark{3}, Akshay K.
Kulkarni\altaffilmark{1}, Min Long\altaffilmark{4}, Richard V.E.
Lovelace\altaffilmark{1}} \altaffiltext{1}{Astronomy Department,
Cornell University, Ithaca, NY 14853} \altaffiltext{2}{Institute for
Mathematical Modeling of the Russian Academy of Sciences, Moscow,
125047, Russia} \altaffiltext{3}{Keldysh Institute of the Applied
Mathematics of the Russian Academy of Sciences, Moscow, 125047,
Russia} \altaffiltext{4}{Center for Theoretical Astrophysics, Dept.
of Physics, University of Illinois at Urbana-Champaign, 1110 W.
Green St., Urbana, IL 61801}

\begin{abstract}
We present results of global 3D MHD simulations of disk accretion to
a rotating star with dipole and more complex magnetic fields using
a Godunov-type code based on the ``cubed sphere" grid developed
earlier in our group. We describe the code and the grid and show
examples of simulation results.
\end{abstract}

\vspace{-0.5cm}
\section{Introduction}

A wide range of stars have significant intrinsic magnetic fields.
Stars are usually strongly magnetized during the protostellar stage (T Tauri stars), and after
collapse to a white dwarf or a neutron star.
Many observational properties
of these stars are determined by the interaction of the accreting
disk matter with the rotating magnetosphere of the star (see e.g.,
Bouvier et al. 2007 for review). In general, the magnetic axis of
the star does not coincide with the rotational axis, due to which
the magnetospheric flow is complicated and the problem requires global
3D MHD simulations. In addition, the magnetic field of the star may
have a complex structure, which adds complications and the necessity to
consider this problem in a global MHD approach.

To solve this problem we developed a special 3D MHD code on the
cubed sphere grid (Koldoba et al. 2002, see also Putman \& Lin 2007)
which is somewhat similar to Yin-Yang grid (Kageyama and
     Sato 2004). This grid has a number of advantages
over spherical or Cartesian grids. A ``cubed sphere"   grid had been
originally developed for the surface of a sphere for geophysical
applications (Sadourny 1972; Ronchi, Iacono, \& Paolucci 1996). In
contrast with these authors, we perform simulations in
three-dimensional space. We used a Godunov-type numerical scheme
(Powell et al. 1999; Kulikowskii, Pogorelov, \& Semenov 2001) and
were able to perform pioneering simulations of disk accretion to
magnetized stars with inclined dipole geometry (Romanova et al.
2003, 2004; Kulkarni \& Romanova 2005). In this paper we show more
recent simulation results obtained with our ``cubed sphere" grid.

%%%%%%%%%%%%%%%%%%%%%%%%%%%%%%%%%%%%%%%%%%%%%
\setcounter{figure}{0}
\begin{figure}[t]
\begin{center}
\includegraphics[scale=0.9]{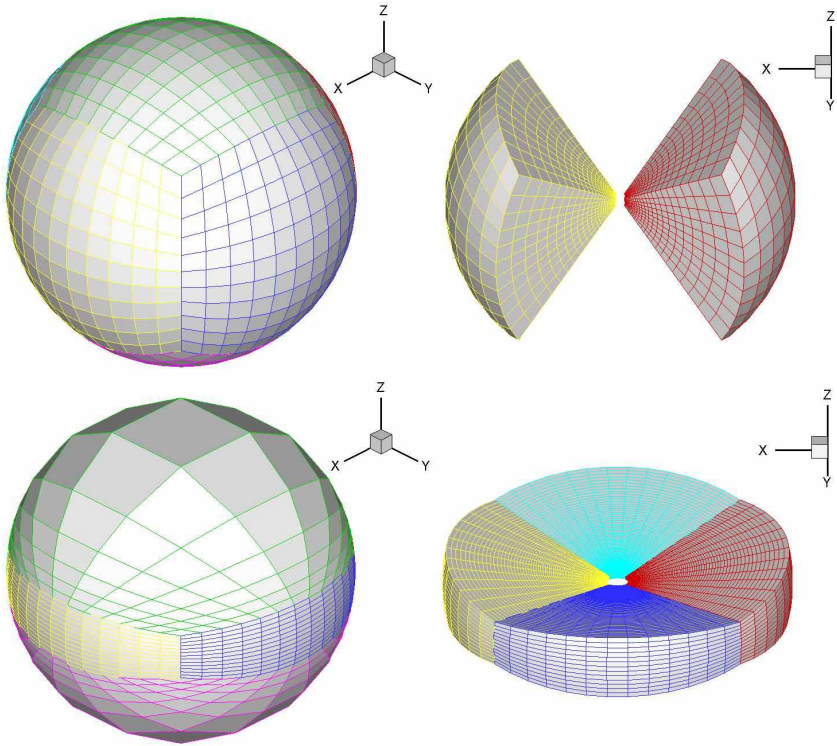}
\end{center}
\caption{{\itshape Top:\/} cubed sphere grid used in simulations.
The grid consists of 6 blocks corresponding to the 6 sides of a cube.
{\itshape Bottom:\/} Cubed sphere grid with high resolution near
the equator.}
\end{figure}

\section{Numerical Method and ``Cubed Sphere" Grid}

We consider disk accretion to a rotating magnetized star.
This problem is difficult to treat
numerically because the  magnetic field varies strongly with
distance from the star ($\sim 1/R^3$ in case of the dipole field and
even more steeply for higher multipoles), and it is rapidly varying in
the laboratory inertial reference frame. To minimize errors in
calculating the magnetic force, the magnetic field $\bf B$ is
decomposed into the ``main'' dipole component of the
star, ${\bf B}_0$, and the
component ${\bf B}_1$ induced by currents in the disk and in the
corona (Tanaka 1994). Another difficulty with this problem is that the
dipole moment changes with time. It rotates with angular velocity
${\bf \Omega}$ so that the ``main" field ${\bf B}_0$ also changes
with time. Consequently, in
the induction equation there is a large term involving ${\bf B}_0$.
To overcome this difficulty we use a coordinate system
 rotating with angular
velocity ${\bf \Omega}$:

$$ {{\partial \rho}/{\partial t}} + {\bf {\nabla}}\cdot (\rho{\bf v})
= 0, $$
 $$ {\partial (\rho{\bf  v})}/{\partial t} + {\bf
{\nabla}}\cdot {T} = \rho {\bf g} + 2\rho ~{\bf v}\times{\bf \Omega}
- \rho~ {\bf \Omega}\times ({\bf\Omega}\times{\bf R}), $$ $$
{\partial (\rho S)}/{\partial t} + {\bf {\nabla}}\cdot (\rho S {\bf
v}) = 0~, $$
$$ {\partial {\bf B}}/{\partial t} = {\bf \nabla
\times} ({\bf v}\times{\bf B}), $$ where  ${\bf v}$ is velocity of
plasma in the rotating frame,  ${\bf B}$ is the magnetic field, and $S$ is the
specific entropy.
 $T$ is the stress tensor with components $T_{ik} \equiv
p\delta_{ik} +\rho v_i v_k+ ({B^2}\delta_{ik}/2-B_iB_k)/4\pi +
\tau_{ik}$. Here $\tau_{ik}$ is viscous stress, $p$ is the gas
pressure. We consider that the viscous stress is determined mainly
by the gradient of the angular velocity because the azimuthal
velocity is the dominant component in the disk. We use the
$\alpha$-viscosity model of Shakura and Sunyaev (1973) with the
coefficient of  dynamic viscosity $\eta_t = \alpha p/\Omega_K$,
where $\alpha$ is a dimensionless coefficient, $\alpha < 1$.   The
viscosity acts only in the accretion disk so that the dominant
contribution to the viscous stress arises from the gradient of the
azimuthal velocity (approximately Keplerian) of the plasma.
  In cylindrical coordinates the non-zero components of the
viscous  stress tensor are
 $$\tau_{r \phi} = \tau_{\phi r} = - \eta_t \frac{\partial{\Omega_K}}{\partial{r}} = \frac{3}{2}p~,$$
where $\Omega_K$ is the Keplerian angular velocity at the given
location. We calculate momentum fluxes due to the viscous stress at
faces of the grid after transforming to the Cartesian coordinates.

\smallskip

\noindent{\it The ``Cubed Sphere" Grid}. The three dimensional grid
consists of a set of concentric spheres of radii $R_j$ in a geometric progression with
$j=1..N_R$.
    The grid on the surface of the
sphere consists of six sectors with the grid on each sector
topologically equivalent to the equidistant grid on the face of a
cube. In each sector the grid of $N\times N$ cells is formed by the
arcs of great circles separated by equal angles. This grid gives
high spatial resolution close to the star which is important for our
study. Recently we incorporated the option to allow the grid to be
compressed towards the equatorial plane, to have higher grid
resolution in the disk (Fig. 1, bottom plots). Such a grid is needed
when higher resolution is required in the disk. Typical grid
resolutions used in our simulations vary from $N_R \times N^2$
=$72\times 31^2$ cells in each of the six sectors, up to
 $N_R
\times N^2$ =$288\times 121^2$ depending on the problem. The cubed
sphere grid naturally lends itself to division into $6\times N$
regions (with $N$ cuts in the radial direction), which are
calculated in parallel using from 48 up to 240 processors.

\smallskip

\noindent{\it Godunov-Type Finite-Difference Scheme}.
    All variables are evaluated at the
centers of the cells, and all vector variables are expressed
in terms of their Cartesian components. Finite difference equations
are written for the Cartesian components of vector variables. The finite
difference scheme of Godunov's type has the form:
$$
{\frac{ {\cal U}^{p+1}-{\cal U}^p}{\Delta t}} V + \sum_{m=1..6} s_m
{\cal F}_m = {\cal Q}~.
$$
  Here,
${\cal U}=\left \{\rho,~ \rho {\bf v},
 {\bf B},
~\rho S\right \} $  is the ``vector" of the densities of conserved
variables; ${\cal F}_m$ is the ``vector" of flux densities normal to
the face ``$m$" of the grid cell, $s_m$ is the area of the face
``$m$", $V$ is the volume of the cell,  $\cal Q$ is the intensity of
sources in the cell, and $\Delta t$ is the time step. To calculate
the flux densities ${\cal F}_m$, an approximate Riemann solver is
used, analogous to the one described by Powell {\it et al.} (1999)
and by  Kulikovskii et al. (2001).

\section{Accretion in stable and unstable regimes}

One of the most striking recent results obtained in our group is the
discovery of accretion through the interchange instability  (Kulkarni \&
Romanova 2008; Romanova et al. 2008). Simulations have shown that a
magnetized star may be either in the stable or unstable regime of
accretion. In the stable regime matter accretes to the star in two ordered
funnel streams and produces ordered hot spots on the surface of the
star, leading to periodic light curves (see Fig. 2,
left panels). In the unstable regime matter penetrates through the
magnetosphere due to the interchange instability forming a small
number (2 to 7) of ``tongues" (see also Li \& Narayan 2004) which
form chaotically at different parts of the inner disk, and the
light-curve from the resulting stochastic hot spots is expected to be irregular (see Fig.2, right
panels).

There are a number of factors which determine the regime of
accretion. If the magnetic axis is inclined with the rotation axis
at a large enough angle, $\Theta>30^\circ$, then the flow is usually
stable (Kulkarni \& Romanova 2009). If the inclination is small,
e.g., $\Theta=5^\circ$ (used in the majority of our simulations),
then the stability of accretion depends on various other factors.

%either stable or unstable accretion is possible depending upon
%a number of other factors.
We compared our simulations with a few relevant theoretical
approaches. The basic theory states that a homogeneous vertical
field at the disk-magnetosphere boundary does not damp azimuthal
perturbations, and therefore is not an obstacle for the development
of unstable $\phi$-modes (e.g. Arons \& Lea 1976). A more general
criterion for magnetized accretion disks states that the disk is
unstable to growth of $\phi$-modes if $\gamma_{B\Sigma}^2 \equiv
-g_{eff} d{\rm ln}(\Sigma/B_z)/dr > 2(r_m d\Omega/dr)^2 \equiv
\gamma_\Omega^2$ (e.g. Spruit, Stehle \& Papaloizou 1995; see also
Kaisig, Tajima \& Lovelace 1992). Here,
$-g_{eff}=r(\Omega_K^2-\Omega^2)$ is the effective gravity, and
$\Sigma= 2 \rho h$ is the surface density. That is, for the
instability to start, the surface density per unit magnetic field
strength $\Sigma/B_z$ should drop off fast enough in the direction
of the star, that the term $\gamma_{B\Sigma}^2$ is larger than the
term associated with the shear, $\gamma_\Omega^2$, which tends to
suppress the instability by smearing out the perturbations. In
addition the term $-g_{eff}$ should be large enough and positive for
the instability to start. In one set of runs we fixed rotation rate
of the star and the initial surface density in the disk  and varied
the accretion rate by varying the $\alpha$-parameter:  $\dot M\sim
\Sigma  v_r\sim \alpha$. We observed that cases with $\alpha < 0.04$
(small $\dot M$) correspond to stable accretion, while cases with
$\alpha > 0.04$ (large $\dot M$) correspond to unstable accretion.
We observed that at larger $\dot M$, the gradient $d{\rm
ln}(\Sigma/B_z)/dr$ is larger and the shear $2(r_m d\Omega/dr)^2$ is
smaller. In addition, at larger $\dot M$ the inner disk comes closer
to the star, which increases $-g_{eff}$. All these factors make
instability more favorable. Thus, we observed that increasing in
accretion rate leads to transition from the stable to the unstable
regime (Kulkarni \& Romanova 2008; Romanova et al. 2008). In another
set of runs we fixed the accretion rate in the disk by fixing
$\alpha$ at some small value, $\alpha=0.02$, but varied the rotation
rate of the star $\Omega$. We observed that at low stellar rotation
rates the flow is unstable, while at higher rates it becomes stable.
This is ascribed to the decrease in the effective gravity,
 $- g _{\rm eff}$, by fast rotation.  The accretion is stabilized
 when the star is spun up and the effective gravity is reduced to a certain point
(Kulkarni \& Romanova 2009). The first effect, namely, the
dependence of the state (stable or unstable) on the accretion rate
$\dot M$, may have important observational consequences: a
particular star may transition between these two regimes and may
thus show intermittency of pulsations. The effect of intermittency
has been observed in a few millisecond pulsars (e.g. Altamirano et
al. 2008). This discovery greatly changes our understanding of
accreting magnetized stars and their possible observational
properties.

%%%%%%%%%%%%%%%%%%%%%%%%%%%%%%%%%%%%%%%%%%%%%
\setcounter{figure}{1}

\begin{figure}[t]
\begin{center}
\includegraphics[scale=0.7]{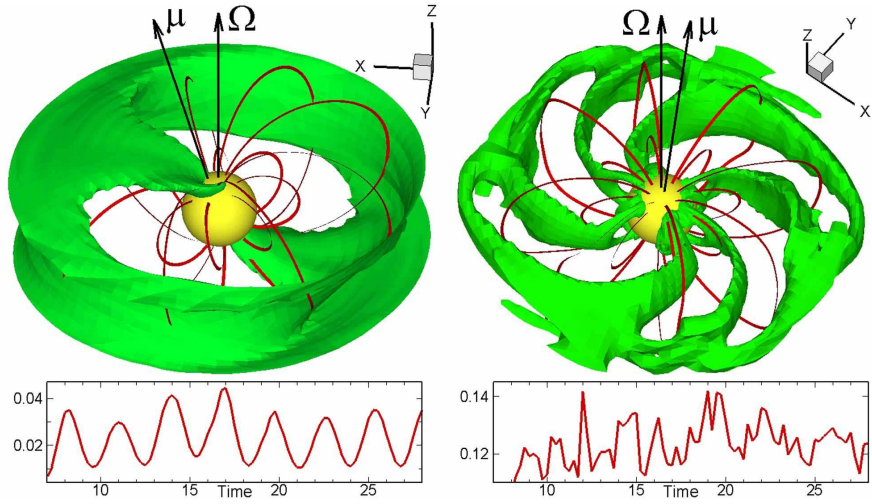}
\end{center}
\caption{{\itshape Left:\/}Accretion in stable regime. The surface
is a constant density surface, and the lines are sample magnetic
field lines. The magnetic axis of the dipole $\mu$ is inclined
relative to the rotational axis $\Omega$ at $\Theta=15^\circ$.
{\itshape Right:\/} Accretion in the unstable regime at
$\Theta=5^\circ$ (from Romanova, Kulkarni \& Lovelace 2008).}
\end{figure}
%%%%%%%%%%%%%%%%%%%%%%%%%%%%%%%%%%%%%%%%%%%%%

\setcounter{figure}{2}
\begin{figure}[!h]
\plottwo{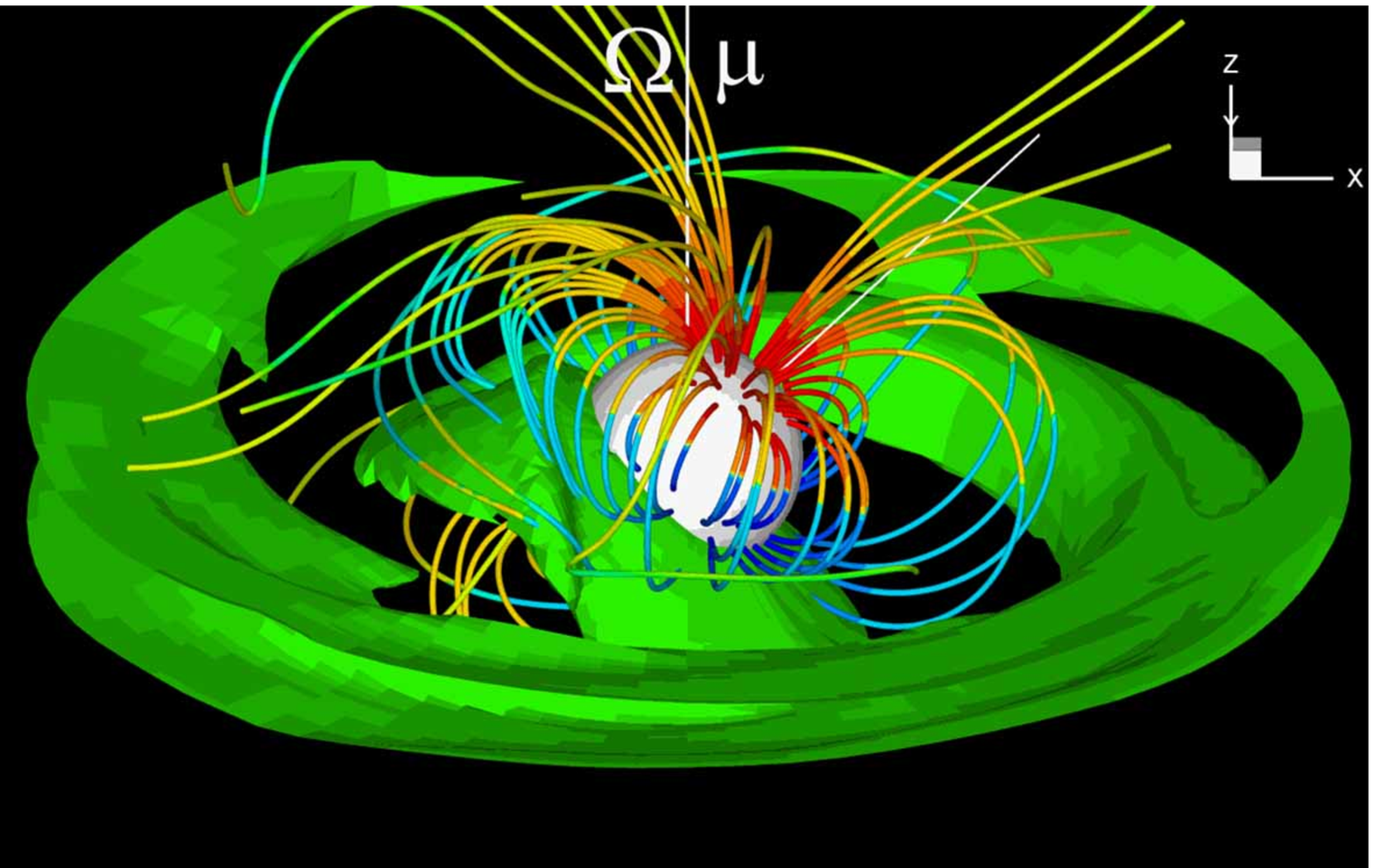}{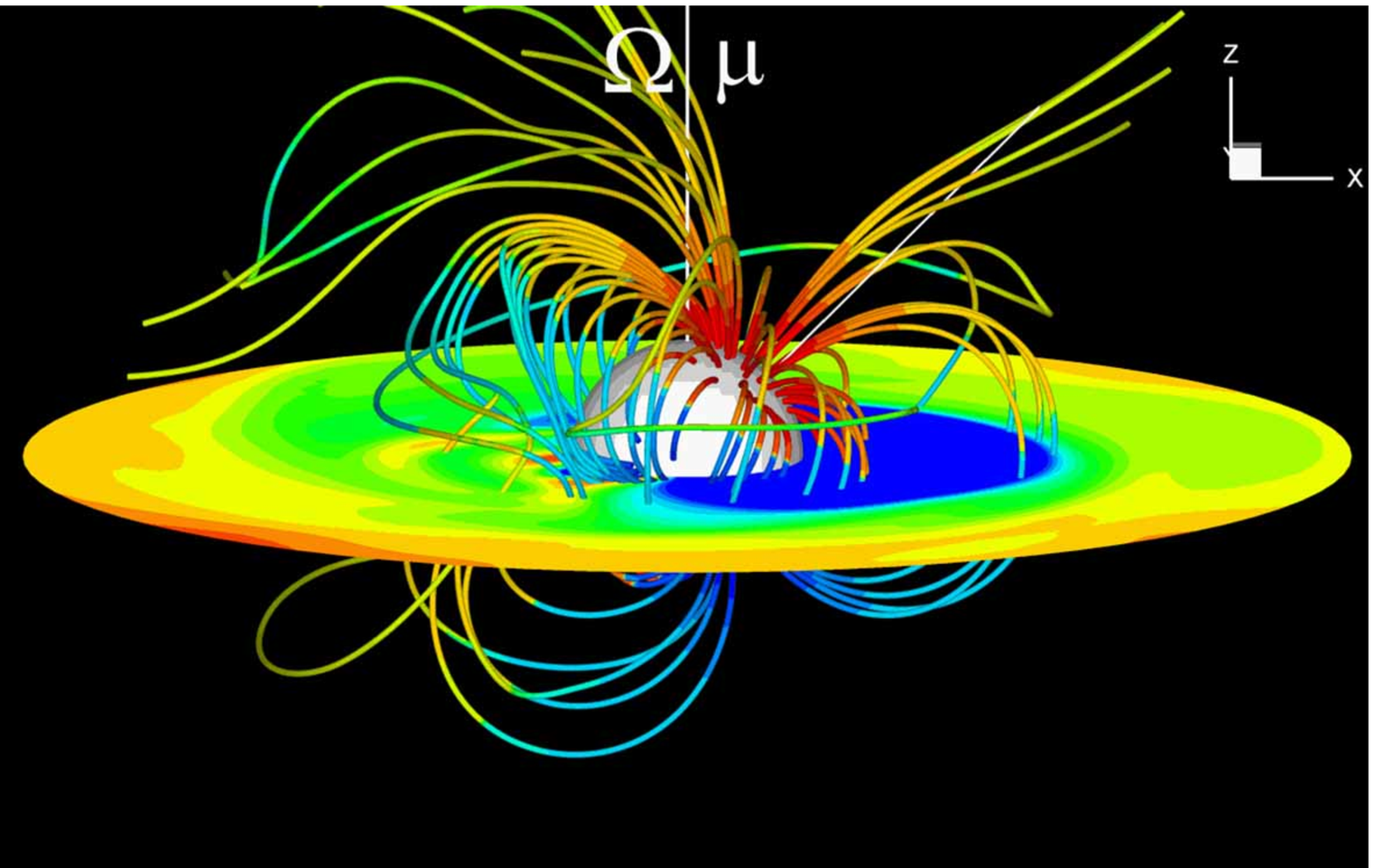} \caption{{\itshape
Left:\/} Accretion to a star with a dipole ($\Theta=45^\circ$) and a
quadrupole ($\Theta_D=30^\circ$) magnetic field (of comparable
amplitudes near the star) misaligned at $\Phi=90^\circ$. A constant
density surface and magnetic field lines are shown. {\itshape
Right:\/} Same as in the left panel, but showing density contours in
the equatorial plane (from Long, Romanova \& Lovelace 2008).}
\end{figure}

\section{Accretion to a star with complex magnetic field}

The stellar magnetic field may be more complex than dipole (e.g.,
Donati et al. 2007). As a first step we investigated accretion to a
star with mixed dipole and quadrupole fields:
$$
\mathbf{B(r)}=\frac{3(\bm{\mu}\cdot{\hat{\bf r}})\hat{\bf
r}-\bm{\mu}}{r^3}+\frac{3D}{4r^4}(5(\hat{\bf D}\cdot\hat{\bf
r})^2-1)\hat{\bf r}-\frac{3D}{2r^4}(\hat{\bf D}\cdot\hat{\bf
r})\hat{\bf D},
$$
where $\bf{\mu}$ is the dipole moment, $\bf{D}$ is the quadrupole
moment, and $\hat{\bf r}$ and $\hat{\bf D}$ are the unit vectors for
the position and the quadrupole moment respectively. In general, the
dipole and quadrupole moments $\bm{\mu}$ and $\bf{D}$ are misaligned
relative to the rotational axis $\bf{\Omega}$, at angles $\Theta$
and $\Theta_D$ respectively. In addition, they can be in different
meridional planes with an angle $\Phi$ between the
$\bf{\Omega}-\bm{\mu}$ and $\bf{\Omega-D}$ planes. First, we
investigated the case when the moments are aligned with each other
but not with the rotational axis. We found that in this case, a
significant amount of matter may flow through the ``quadrupole
belt'' forming a ring-shaped spot on the surface of the star (Long
et al. 2007). In the more general case when the dipole and
quadrupole moments are not in the same plane, the field is more
complicated with a number of poles of different polarity on the
surface of the star (Long et al. 2008). The accreting matter chooses
the most energetically favorable path, due to which the funnel
streams are often quite close to the equatorial plane (see Fig. 3).
Work on accretion to a star with a higher multipolar field is in
progress.

\acknowledgements The authors were supported in part by NASA grant
NNX08AH25G and by NSF grants AST-0607135 and AST-0807129. MMR is
thankful to NASA for using NASA High Performance Facilities. AVK and
GVU were supported in part by grant RFBR 06-02016608, Program 4 of
RAS. MMR and RVEL thank the organizers for a very interesting meeting.

\end{document}